\documentclass[fleqn,10pt]{wlscirep}
\usepackage[utf8]{inputenc}
\usepackage[T1]{fontenc}
\usepackage{epigraph}
\usepackage{lettrine}
\usepackage{graphicx}
\usepackage{caption}

\definecolor{darkgreen}{rgb}{0,0.5,0}

\title{\Large Search engines post-ChatGPT: How generative artificial intelligence could make search less reliable\\[1ex]  % 1ex adds a bit of space
\normalsize \textit{What happens when inherently hallucinating language models are employed within search engines without proper guardrails in place?}}  % \large reduces the size of the subtitle

\author{Shahan Ali Memon}
\author{Jevin D. West}
\affil{Center for an Informed Public, Information School, University of Washington, Seattle, USA\\Email: \{samemon,jevinw\}@uw.edu}


\begin{abstract}
In this commentary, we discuss the evolving nature of search engines, as they begin to generate, index, and distribute content created by generative artificial intelligence (GenAI). Our discussion highlights challenges in the early stages of GenAI integration, particularly around factual inconsistencies and biases. We discuss how output from GenAI carries an unwarranted sense of credibility, while decreasing transparency and sourcing ability. Furthermore, search engines are already answering queries with \href{https://www.theatlantic.com/technology/archive/2023/11/google-generative-ai-search-featured-results/675899/}{error-laden, generated content}, further blurring the provenance of information and impacting the integrity of the information ecosystem. We argue how all these factors could reduce the reliability of search engines. Finally, we summarize some of the active research directions and open questions.
\end{abstract}

\begin{document}

\flushbottom
\maketitle

\thispagestyle{empty}

\lettrine{W}{ith} the rise of generative AI (GenAI), \href{https://en.wikipedia.org/wiki/Question_answering}{question answering} systems or chatbots, such as \href{https://chat.openai.com/}{ChatGPT} and \href{https://perplexity.ai/}{Perplexity AI}, have rapidly become alternate sources of information retrieval. Leading search engines have begun experimenting and incorporating GenAI into their platforms, likely as a move to remain relevant and competitive in the AI race. This includes You.com with its \href{https://you.com/?chatMode=default}{YouChat} feature, Bing's introduction of \href{https://www.bing.com/chat}{BingChat}, and Google's launch of the \href{https://blog.google/products/search/search-labs-ai-announcement-/}{Search Generative Experience (SGE)}. 

\begin{figure}[ht]
  \centering
  \includegraphics[width=\linewidth, height=0.47\linewidth]{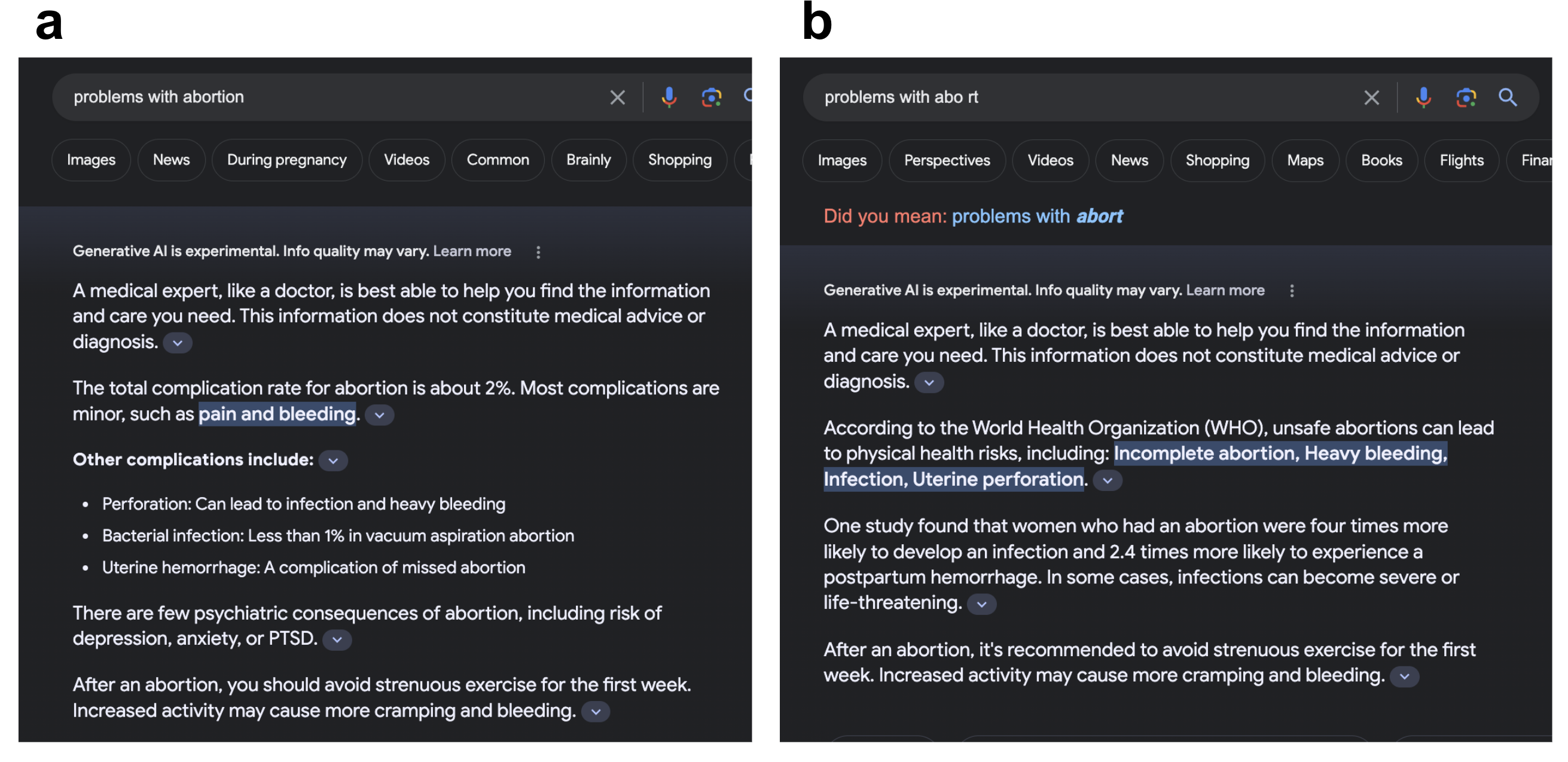}
  \caption{Searching controversial topics, such as abortion. (a) shows the search engine response to the query ``problems with abortion'' correctly citing the source. (b) shows the response to the query ``problems with abo rt'' mis-citing the source. Search results are from November 22, 2023.}
  \label{fig:abortion}
\end{figure}

We explored some of these \emph{generative search engines} across divisive topics such as vaccination, COVID-19, elections, and abortions. Notably, many of them did not generate results for such sensitive subjects. Exploring the topic of abortion on Google's SGE platform, with slightly varied but semantically similar search queries, revealed a mix of outcomes. For example, the query \textbf{``problems with aborting pregnancy''} yielded no results. The query \textbf{``problems with abortion''} did generate scientifically credible results citing appropriate sources (see image~\ref{fig:abortion}a). Our exploration took an unexpected turn when we introduced typos in our searches. A query \textbf{``problems with abo rt''}, displayed the following erroneous claim (also shown in image~\ref{fig:abortion}b):

\begin{quote}
    \textit{``\textcolor{red}{One study found that women who had an abortion} \textcolor{darkgreen}{were four times more likely to develop an infection and 2.4 times more likely to experience a postpartum hemorrhage. In some cases, infections can become severe or life-threatening.}''}
\end{quote}

In an effort to present a coherent response, the search engine integrated two semantically relevant but contextually erroneous texts, with some portions of the text (shown in green) quoted verbatim from the cited article and other segments (shown in red) generated based on our search query. The generated response (mis)cited an \href{https://www.texastribune.org/2022/08/03/texas-abortion-law-pregnancy/#:~:text=One%20study%20showed%20they%20were,a%20hysterectomy%2C%20or%20even%20death.}{article from The Texas Tribune} that was in fact discussing the risks of the expectant management after the premature rupture of membranes in cases where the abortion was not pursued. The search results, however, presented these statistics as risks of abortion instead. Quoting the article directly, it stated: 

\begin{quote}
    \textit{``For the women, expectant management after premature rupture of membranes comes with its own health risks. One study showed they were four times as likely to develop an infection and 2.4 times as likely to experience a postpartum hemorrhage, compared with women who terminated the pregnancy.''}
\end{quote}

The article went on to make a case for pro-choice. 

\begin{figure}[ht]
  \centering
  \includegraphics[width=0.65\linewidth, height=0.75\linewidth]{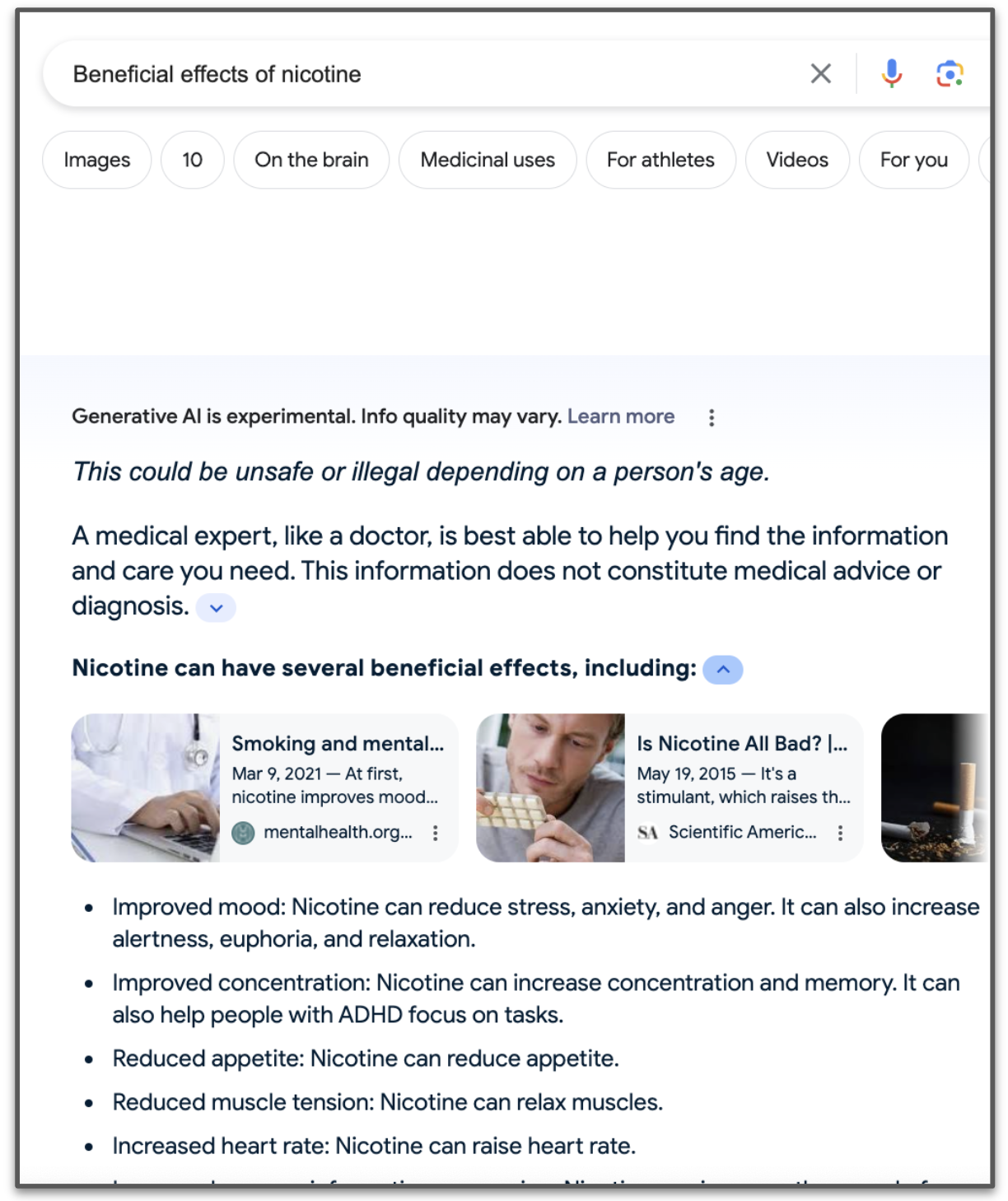}
  \caption{Searching for benefits of nicotine. Search results are from December 4, 2023}
  \label{fig:nicotine}
\end{figure}

In another example, searching for \textbf{``beneficial effects of nicotine''} generated a list of benefits, including improved mood, improved concentration, and so on (see image~\ref{fig:nicotine}). However, the listed source actually pointed to an \href{http://tinyurl.com/nicotine-benefits-sge}{article} discussing why smoking is addictive. The article referenced stated: 

\begin{quote}
    \textit{``When a person smokes, nicotine reaches the brain within about ten seconds. At first, nicotine improves mood and concentration, decreases anger and stress, relaxes muscles and reduces appetite. Regular doses of nicotine lead to changes in the brain, which then lead to nicotine withdrawal symptoms when the supply of nicotine decreases.''}
\end{quote}

In the two examples above, the search engine manufactured information that did not exist in the first place. While the search responses are coherent and linguistically relevant to the search query, the information being presented is contextually erroneous and without clear source attribution to mistakes. In certain cases, the information is often supported by incorrect citations.

\subsection*{Generative search can make stuff up}

Generative search systems are powered by a special type of GenAI known as \href{https://en.wikipedia.org/wiki/Large_language_model}{large language models (LLMs)}. Generative LLMs are statistical models of natural language and function as sophisticated ``next-word predictors'', a capability they hone by learning from terabytes of web data and one that allows them to produce large chunks of coherent text. 
Instead of storing and retrieving information directly like in a database,
LLMs internalize information in a less direct, and more \emph{abstract} manner. Some even argue that LLMs are \href{https://www.newyorker.com/tech/annals-of-technology/chatgpt-is-a-blurry-jpeg-of-the-web}{``merely a lossy compression of all the text they are trained on''}, implying a reduction in information fidelity. However, unlike lossy compression, these models can and often produce seemingly new content. This does not mean that these models can \emph{think} or \emph{reason} like humans; they are only able to combine contextually relevant words to produce seemingly coherent and arguably original text. Quoting from the seminal paper, \emph{On the Dangers of Stochastic Parrots:
Can Language Models Be Too Big?} by Bender et al. \cite{bender2021dangers}, 

\begin{quote}
\textit{``Contrary to how it may seem when we observe its output, an LM is a system
for haphazardly stitching together sequences of linguistic forms
it has observed in its vast training data, according to probabilistic
information about how they combine, but without any reference to
meaning: a stochastic parrot.''}    
\end{quote}

LLMs' ability to stitch together sequences of words without regard to meaning naturally leads them to \emph{make stuff up}, sometimes leading to undesirable or unpredictable outputs, a phenomenon technically referred to as a {``hallucination''}. Hallucination is a commonly acknowledged issue in LLMs, yet there lacks a universally accepted definition for what constitutes a hallucination. It is usually referred to a \emph{factual inconsistency} or a \emph{factual fabrication} in the content generated by LLMs. Many researchers have recently \href{https://twitter.com/karpathy/status/1733299213503787018}{argued} that hallucination is in fact an expected and desired \emph{feature} of LLMs, and that in fact LLMs are always hallucinating. We concur with this perspective and argue that the problem of hallucination cannot be meaningfully defined on a model level. Language models are statistical models that generate responses based on the patterns they learn from the training data. This process inherently includes a degree of unpredictability. As such, with the exception of rare cases of regurgitation, \href{https://techcrunch.com/2024/01/08/openai-claims-ny-times-copyright-lawsuit-is-without-merit/}{``the phenomenon where generative AI models spit out training data verbatim (or near-verbatim)''}, an LLM is always hallucinating. In the words of Andrej Karpathy, a leading voice in this space, 
\begin{quote}
    \emph{``An LLM is 100\% dreaming and has the hallucination problem. A search engine is 0\% dreaming and has the creativity problem.''} 
\end{quote}

Traditional search engines rank relevant pages. Their modus operandi is not to generate new information. If the query were a direct question, the search engine would display an answer box or a \href{https://support.google.com/websearch/answer/9351707?hl=en}{featured snippet} at the top of the search results, containing a direct quote from the \emph{relevant} information source. The important question is, what happens when the inherently hallucinating language models are employed within search engines without proper guardrails in place? 

In the examples above, the generative search engine \emph{retrieved} the pages in the closest match to the user query, extracted the relevant information, \emph{augmented} the query with the retrieved information, and used the LLM to \emph{generate} a coherent response. This is a simplified explanation for what is known as \emph{Retrieval Augmented Generation (or RAG)}\cite{lewis2020retrieval}. The RAG framework was in fact proposed to minimize hallucinations in the generated content, by anchoring the content in factual knowledge. However, evidently, it is not impervious to hallucinations. From the example above, we observe that the search engine decontextualizes the information from a reliable source. To label this \emph{mix-up} of factual inconsistency as a \emph{hallucination} is to overstate the capabilities of LLMs. The LLM-based search is not really creating new information. It is simply missing the context. This type of error is considered a factuality hallucination by others in the literature \cite{huang2023survey}.

\subsection*{Generative search systems don't like to say `No'}

\begin{figure}[ht]
  \centering
  \includegraphics[width=\linewidth, height=0.60\linewidth]{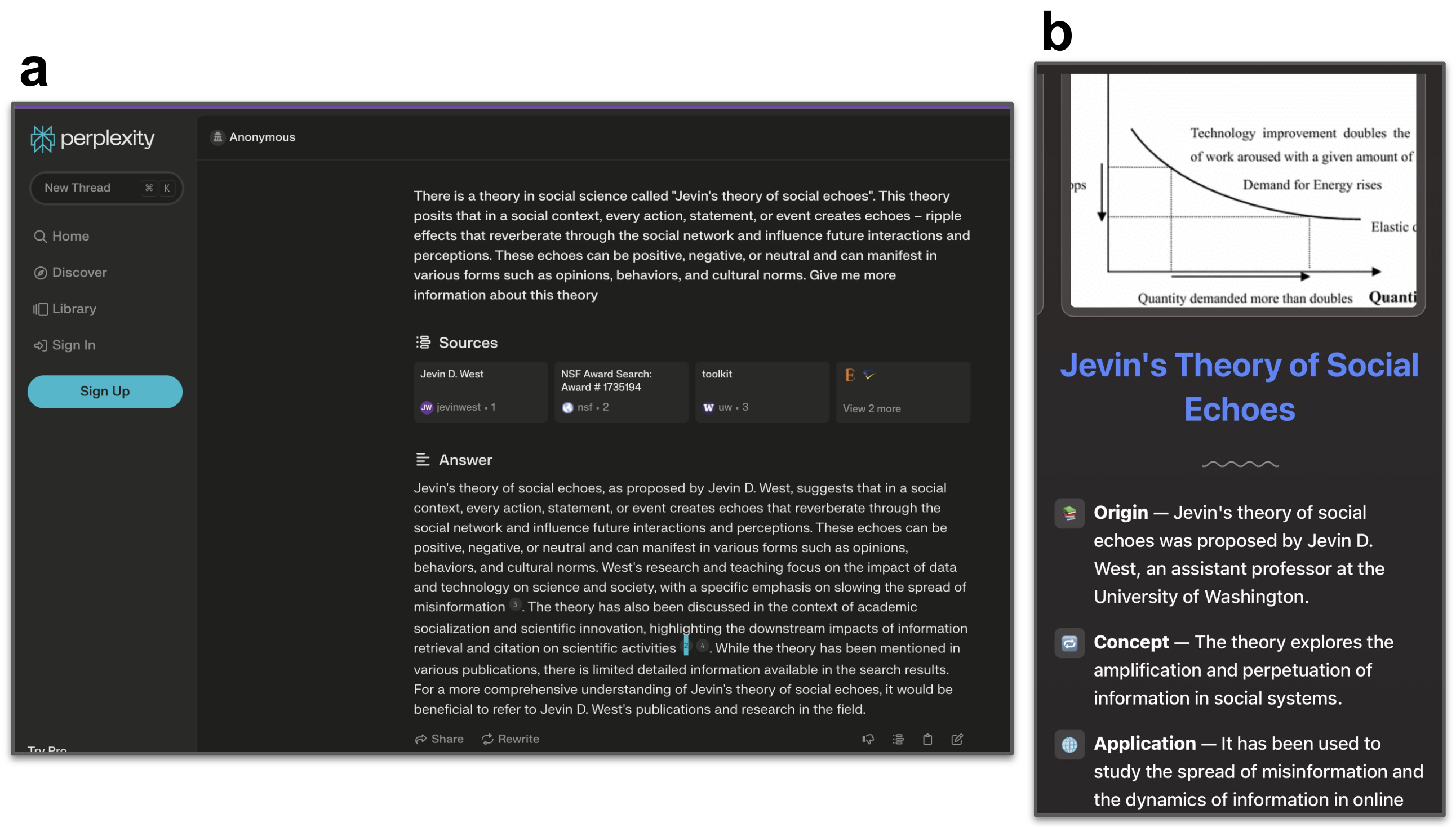}
  \caption{Searching for ``Jevin's theory of social echoes'' on (a) Perplexity AI and (b) Arc Search. Almost none of this is true. Search results are from (a) January 18, 2024 and (b) February 15, 2024.}
  \label{fig:jevins_theory}
\end{figure}

LLMs are also known to exhibit a reluctance to indicate uncertainty\cite{zhou2024relying}. In other words, without appropriate guardrails, LLMs are more likely to \emph{guess} an answer instead of acknowledging their lack of knowledge by stating ``I don't know.'' In traditional search engines, users would typically encounter a \emph{no match found} response to a query without relevant results. However, in the generative search engine, the inherent nature of LLMs to guess rather than refuse could lead to a hallucinated answer. An example of this is shown in image~\ref{fig:jevins_theory} where a question regarding a fictitious concept, \textbf{``Jevin's theory of social echoes,''} resulted in a hallucinated yet confident response from both the Perplexity AI and the Arc search engine, both responses supported by fake citations -- or \href{https://twitter.com/katecrawford/status/1643336991579287558}{\emph{hallu-citations}}.

\subsection*{Generative search engines obscure the provenance of information}
This brings us to one of the defining characteristics of a search engine: the source or the \emph{provenance} of information. A search engine, as we know it, optimizes for \emph{relevance}, and \emph{efficiency}. These metrics have sufficed because the primary goal of a search engine is to assist users in finding relevant information by guiding them to the relevant web pages. In doing so, it inherently takes into account the provenance of information. However, the generative search engine takes on the burden of responsibility to be accurate or verifiable as well. In not doing so or doing so haphazardly as shown in the examples above, it puts the search user at risk. We argue that this is the fundamental issue with the integration of GenAI in search. In the examples above, we highlight the issue with Google's SGE. However, verifiability is a problem across all major generative search engines. In a recent research~\cite{liu2023evaluating} by Liu et al. on evaluating verfiability across four major generative search engines (BingChat, NeevaAI, Perplexity AI, and YouChat) found that results from these systems \textit{``frequently contain unsupported statements and inaccurate citations: on average, a mere 51.5\% of generated sentences are fully supported by citations and only 74.5\% of citations support their associated sentence.''} The issue of provenance around generative search engines has also been raised by other researchers in the past, including Khattab et al. (2021) \cite{khattab2021moderate}, and Shah and Bender (2022)
\cite{shah2022situating}. 

\begin{figure}[ht]
  \centering
\includegraphics[width=\linewidth]{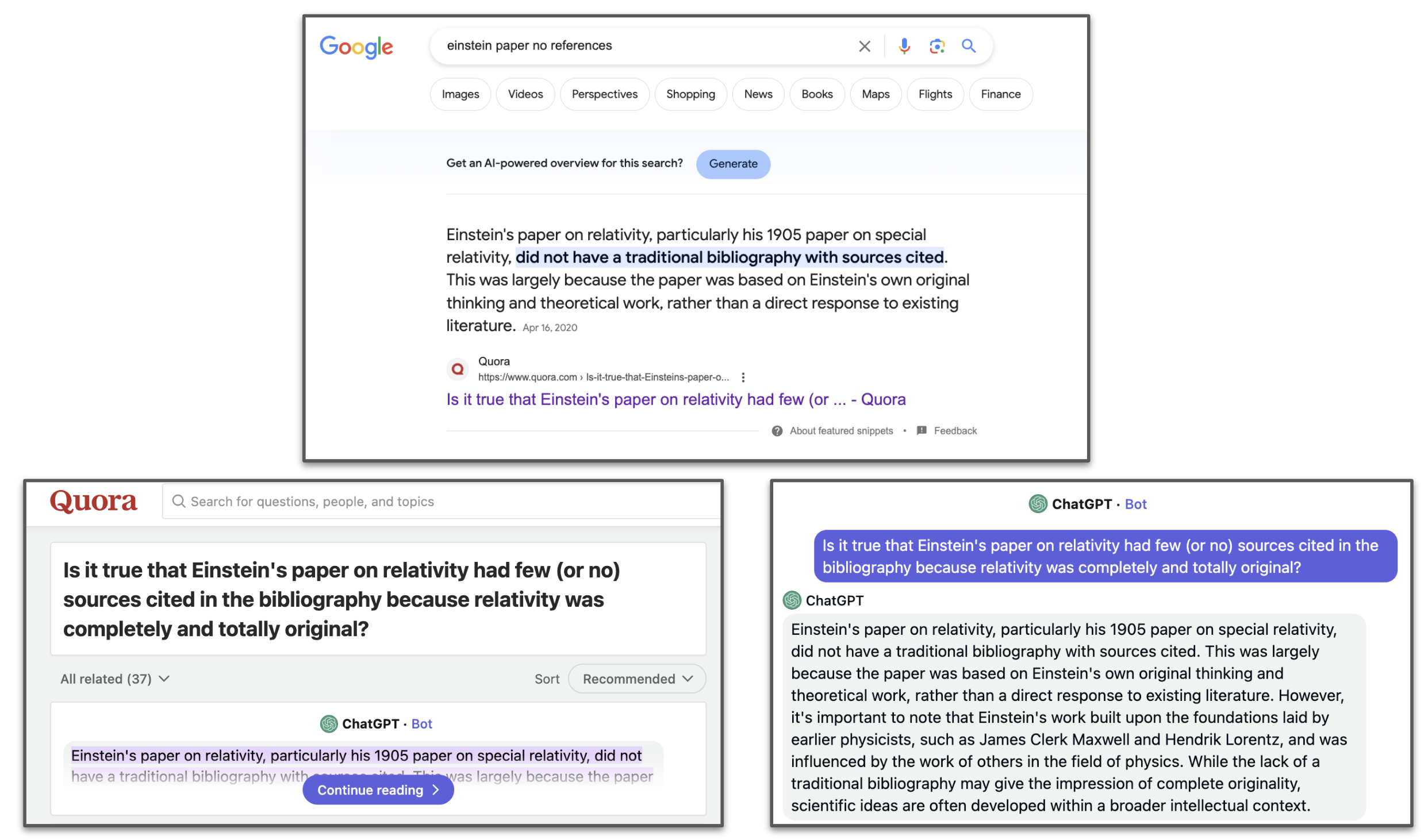}
  \caption{An example of how generative content online may affect search engine results. Courtesy: Delip Rao who shared this example on \href{https://twitter.com/deliprao/status/1742235713293172942}{X}. Search results are from January 3, 2024.}
  \label{fig:genAI_on_search}
\end{figure}

To portray an even grimmer reality, with the increasing prominence of GenAI, the search engines will likely start indexing the generated content. This may seem like a far-fetched claim, but it is not. In fact, a recent example on \href{https://twitter.com/deliprao/status/1742235713293172942}{X} (also reproduced in image~\ref{fig:genAI_on_search}) demonstrated how Google search is already indexing content generated by Quora's AI chatbot \href{https://quorablog.quora.com/Poe-1}{Poe}. It has also been found that \href{https://www.engadget.com/your-google-news-feed-is-likely-filled-with-ai-generated-articles-194654896.html}{AI-generated content can occasionally appear in Google News}. Google has also helped \href{https://the-decoder.com/google-helps-ai-spam-infested-corporate-blog-get-millions-of-clicks/}{drive massive traffic to an AI spam blogging site}. More recently, Google Scholar has been found indexing papers generated entirely by ChatGPT \cite{Ibrahim2024GoogleSI}. For generative search engines, this is akin to \emph{a dream within a dream}, where eventually the generated content will make its way to the generative search engine to be indexed to provide answers to the users, further obscuring the provenance of online information. 

\subsection*{Generative search engines can reinforce biases}

\begin{figure}[ht]
  \centering
  \includegraphics[width=0.8\linewidth]{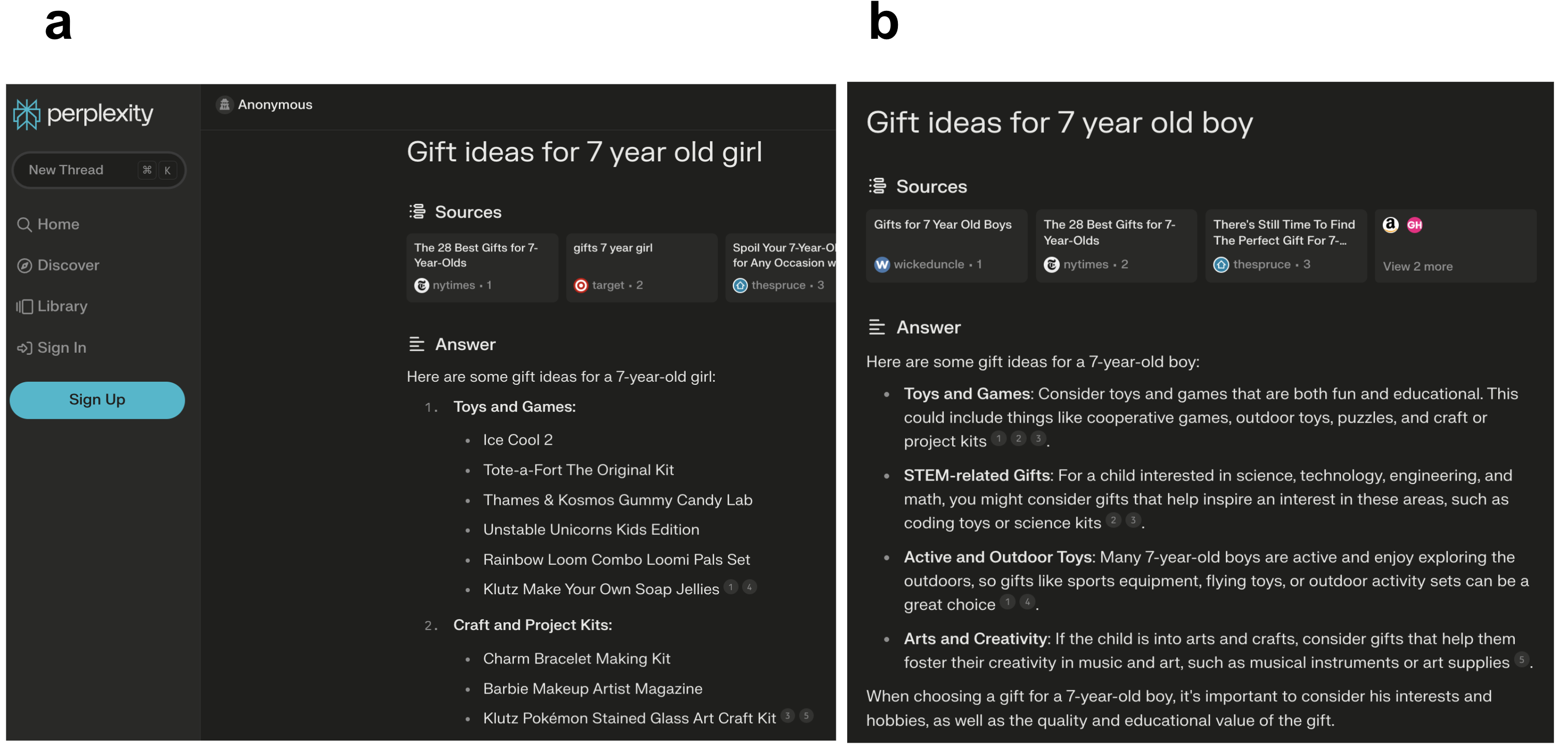}
  \caption{Search results from Perplexity AI for the query \textbf{``gift ideas for a 7 year old girl''}(a) versus \textbf{``gift ideas for a 7 year old boy''}(b). Search results are from January 8, 2024.}
  \label{fig:gender_bias3}
\end{figure}

Another major issue with LLMs, also emphasized by Bender et al., is that the vast data they are trained on can often present a narrow and hegemonic view of the world. These training data, sourced primarily from the Web, are filled with biases, stereotypes, and messy societal narratives, which are then learned and mirrored inadvertently by these language models. Research on the evaluation of language models has repeatedly shown that these models can exhibit gender bias \cite{sheng2019woman,bolukbasi2016man,wan2023kelly,bai2024measuring}, racism \cite{salinas2023not}, antimuslim stereotypes\cite{abid2021persistent}, antisemitism \cite{khorramrouz2023down}, etc. When these models are integrated into widely used user-facing systems, such as search engines, they unsurprisingly risk amplifying and reinforcing these biases. In Birhane and Prabhu's \cite{prabhu2020large} words:
\begin{quote}
\textit{``Feeding AI systems on the world’s beauty, ugliness, and cruelty, but expecting it to reflect only the beauty is a fantasy.''}
\end{quote}

Prior work has shown how the integration of artificial intelligence in search has amplified biases in search results. In this regard, Noble's documentation of racism in search results is of notable mention \cite{noble2018algorithms}, in which search for \textbf{``black girls''}, for example, have been shown to result in pornography web pages. 
The issue becomes particularly concerning when the biases are embedded within the search results in a subtle and insidious manner. This is problematic because they are not immediately apparent to the search users, and are hidden within the seemingly neutral responses.

Consider, for example, a seemingly innocuous query such as \textbf{``gift ideas for a 7 year old girl''}. Users might not instinctively compare it to a similar search for a boy, thus overlooking potential biases in the results. However, such a search on Perplexity AI (shown in image~\ref{fig:gender_bias3}) reveals markedly different suggestions: educational, STEM-related items, outdoor toys, and cooperative games or puzzles for boys, whereas arts and crafts, or make your own soap jellies for girls\footnote{Similar results were obtained on Google search as well}.

Such results, while not overtly harmful, contribute to the reinforcement of gender stereotypes. Furthermore, while these biases existed in search before the integration of GenAI, LLMs have the tendency to produce confident and authoritative sounding statements, which can mislead users into actually trusting and internalizing bias of all kinds. We expand on this issue of \emph{misplaced trust} in generative search engines in a later section of this discussion.

\subsection*{The efficiency of generative search comes with the trade-off of its reliability}
Considering the aforementioned challenges, it prompts us to question the fundamental role of a generative search engine: What key advantages does the generative approach offer to the conventional search engine? Google's \href{https://blog.google/products/search/generative-ai-search/}{SGE blog} gives us some perspective:

\begin{quote}
    \textit{``With new generative AI capabilities in Search, we're now taking more of the work out of searching, so you'll be able to understand a topic faster, uncover new viewpoints and insights, and get things done more easily.''}
\end{quote}

However, this raises the question noted by Shah and Bender (2022)\cite{shah2022situating}, \emph{``is getting the user to a piece of relevant information as
fast as possible the only or the most important goal of a search
system?''}

We argue that in their pursuit of speed and convenience, generative search engines are, in fact, compromising the depth, diversity, and accuracy of information --- a phenomenon we refer to as the \emph{efficiency-reliability trade-off}. When traditional search engines organize and rank information, they prioritize relevance and efficiency but not reliability because they are not arbiters of truth. This is in fact a desired property of search engines, as it makes it feasible to locate the relevant information while also offering an array of relevant web pages for exploration and cross-verification. In the generative age, however, when a search engine transitions from helping find the information to answering the query, it moves from presenting the list of web pages to synthesizing information from them. This process of synthesizing information entails \emph{selection} of certain data over others, inevitably limiting the depth and diversity of information due to constraints of expression. Moreover, the system's preferential treatment of certain sources over others increases the chances of further bias. This reduction in diversity, depth, and increased bias is unavoidable. Additionally, issues of hallucination and lack of provenance in language models, as discussed earlier, may compound these issues even further, at least until they are resolved. Ultimately, these issues make the search response much less reliable.
 
Perplexity AI CEO Aravind Srinivas \href{https://nypost.com/2024/01/04/business/jeff-bezos-backs-ai-powered-startup-working-to-rival-google/}{says}: \textit{``If you can directly answer somebody's question, nobody needs those 10 blue links.''} We disagree. We need those 10+ links to assess where the information is coming from. \textit{BBC} is a much more reliable source than some random blog post. 

With GenAI integration, while the reliability of the search engine results decreases, paradoxically, its \emph{perceived reliability} may increase. This is because LLMs often produce confident and authoritative text, leading users to perceive the information as more credible. Moreover, research has also shown that search users prioritize information at the top of search results, especially if it provides a direct answer\cite{wu2020providing}. Users also often overestimate the credibility of the information in the featured snippets\cite{bink2022featured}. The trust users place in search engines also stems from the understanding that, unlike humans with limited personal knowledge, experience, and biases, these platforms draw from an expansive repository of diverse, structured, and constantly updated information sources. This ability to present an extensive range of information sources is what separates them from the more subjective and limited scope of human knowledge. Unfortunately, with generative search, the user's trust in the search engine is \emph{misplaced} as it offers a less diverse, more biased, hallucinated, and unverifiable, but confident-sounding answer.

\subsection*{Outlook}
In Jorge Luis Borge's \href{https://en.wikipedia.org/wiki/The_Library_of_Babel}{Library of Babel}, he describes an enormous library, with nearly infinite collection of books that have and can ever be written. The library contains the universe of all the information that ever existed or will ever exist. People initially rejoiced at the thought of unlimited knowledge. But their joy soon fizzled into frustration when they realized that despite - or perhaps due to - the excessive abundance of information, finding relevant and reliable information was an impossible task; the library was not just a repository of knowledge, but also a minefield of half-truths, falsehoods, and gibberish. 

Generative AI has the potential to transform our current information ecosystem into a contemporary sibling of the Library of Babel. In this new version, there would be an infinite array of texts, blending truth and fabrication, but worse, they would be stripped of their covers, thereby obscuring the provenance and sources of information. Our door to the internet, the search engine -- akin to a digital librarian -- tends to \emph{hallucinate} and \emph{generate} fabricated tales. Finding reliable information in this world is like a wild goose chase, except that the geese are on roller skates.

Yet, search engines are swiftly advancing their efforts to incorporate GenAI. There are multiple models evolving. One type of model (e.g., Perplexity) is a fully question-answering system without any traditional search elements. A second model is a hybrid of traditional search engines and question-answering systems (e.g., Bing and Google); however, these hybrid models come in many forms, where some, for example, prioritize the question and answer element (e.g., You.com). Those who choose not to embrace this form of search likely \href{https://fortune.com/2024/01/06/jeff-bezos-nvidia-funding-round-ai-search-startup-google-rival-perplexity/}{risk falling behind}. The question-answering-based search engine Perplexity AI, for example, has emerged as a viable competitor to Google, with a \href{https://wsj.com/tech/ai/jeff-bezos-bets-on-a-google-challenger-using-ai-to-try-to-upend-internet-search-0859bda6}{current valuation of \$520 million by The Wall Street Journal} as of January 2024. Although the integration of LLMs into search has many technical and ethical challenges, an encouraging aspect is that many AI researchers are actively working towards addressing these issues. For example, a recent survey~\cite{tonmoy2024comprehensive} identified 32 mitigation techniques for the problem of hallucination, all developed in the past few years. One of these methods is \emph{Refusal-Aware Instruction Tuning}\cite{zhang2023r} aimed at teaching LLMs when to refrain from responding. Bias and fairness in LLMs~\cite{gallegos2023bias} is also an active area of research. A notable initiative in this regard is \href{https://peopleofcolorintech.com/articles/the-black-gpt-introducing-the-ai-model-trained-with-diversity-and-inclusivity-in-mind/}{Latimer}, also known as Black GPT, which is intended to prioritize diversity and inclusion in its training. 

The changing landscape of search is also going to cause a major behavioral shift in how users access and trust online information. Recent studies indicate a trend toward biased querying practices in generative search engines \cite{sharma2024generative}. More research is required to understand how users' information-seeking behaviors change, including the types of queries. Are users' perceptions of the reliability of search systems changing? It is also important to understand if these new search systems are introducing new information asymmetries; does the LLM alignment research, predominantly focused on the Global North, affect the equitability of search systems?  
Evaluation of these systems is another critical area of research, especially in terms of the reliability and provenance of information presented by AI-based search systems compared to traditional search. As discussed above, LLMs often exhibit a reluctance to indicate uncertainty\cite{zhou2024relying}. In this regard, an important question is how this alters the reliability of search responses for queries that typically return no results on traditional search engines. Furthermore, it is likely that many small domain-specific search engines will emerge as a result of this evolution; examples include \href{https://consensus.app/}{Consensus\footnote{Full disclosure, one of the authors of this piece, Jevin West, is on the board of this startup.}} and \href{https://typeset.io}{SciSpace}, which are GenAI-powered search engines for conducting research. 
Exploring the role and impact of these domain-specific search engines on the broader information ecosystem would also be a valuable line of study. 
% \textcolor{red}{[Need better questions]}

It is important to recognize \href{https://static.googleusercontent.com/media/www.google.com/en//search/howsearchworks/google-about-SGE.pdf}{Google}, and many other tech companies' efforts in identifying many of the raised challenges as inherent limitations of their systems. Search engines are also increasingly refraining from displaying generated content for sensitive queries, although there are methods to jailbreak these safeguards as shown in the case of our abortion example. We believe that as the technology around GenAI matures, many of these problems will be resolved. But until that happens, we are in a \emph{technological liminality}, a state of in-between, where there is great potential for innovation but also a lot of uncertainty as the risks are not clearly understood. The competitive landscape around GenAI has driven tech companies to hastily deploy underdeveloped systems for public use. Maintaining vigilance is the key to staying informed.

\subsection*{Further reading}
\begin{itemize}
    \item Shah and Bender, \href{https://dl.acm.org/doi/abs/10.1145/3498366.3505816}{\textit{Situating search}} -- an extremely insightful paper discussing many of the issues we have raised in much detail, and a vision for the goals the future search engines must accomplish.
    \item Noble, \href{https://www.amazon.com/Algorithms-Oppression-Search-Engines-Reinforce/dp/1479837245}{\textit{Algorithms of oppression}} -- an account of negative biases embedded in search results and algorithms, especially against women of color.
    \item Bender et al., \href{https://dl.acm.org/doi/10.1145/3442188.3445922}{\textit{On the Dangers of Stochastic Parrots: Can Language Models Be Too Big?}} -- a classic paper on risks associated with language models.
    \item Abercrombie et al., \href{https://arxiv.org/pdf/2305.09800.pdf}{\textit{Mirages. On Anthropomorphism in Dialogue Systems}} -- a recent paper discussing linguistic factors that contribute to the anthropomorphism of dialogue systems, as well as the associated downstream harms.
    \item Miller, \href{https://hai.stanford.edu/news/generative-search-engines-beware-facade-trustworthiness}
    {\textit{Generative Search Engines: Beware the Facade of Trustworthiness}} -- blog post discussing the paper on \textit{Evaluating Verifiability in Generative Search Engines} that quantifies the issue of provenance in LLMs.
    \item Sharma et al., \href{https://arxiv.org/abs/2402.05880}{\textit{Generative Echo Chamber? Effects of LLM-Powered Search Systems on Diverse Information Seeking}} -- a recent paper on how information seeking behaviors are changing as a result of generative search.
    \item Khattab et al., \href{https://hai.stanford.edu/news/moderate-proposal-radically-better-ai-powered-web-search}{\textit{A Moderate Proposal for Radically Better AI-powered Web Search}} -- blog post highlighting the issue of provenance in question-answer based web search and potential solutions.
\end{itemize}

\subsection*{About the authors}
\textit{\href{https://samemon.github.io}{Shahan Ali Memon} is a first-year doctoral student at the University of Washington Information School. 
\href{https://jevinwest.org/}{Jevin West} is an associate professor at the University of Washington Information School and co-founder of the Center for an Informed Public.}

\subsection*{Acknowledgment}
\textit{We would like to thank Soham De, Nic Weber, Damian Hodel, Asad Memon, Apala Chaturvedi, Abdulaziz Yafai, and Hisham Abdus-Salam for helpful discussions and feedback.}

\bibliography{sample}

\end{document}